# A MALWARE CLASSIFICATION SURVEY ON ADVERSARIAL ATTACKS AND DEFENSES

Ponnuru Mahesh Datta Sai[1], Likhitha Amasala[2], Tanu Sree Bhimavarapu[3], Guna Chaitanya Garikipati[4]

mp1366@srmist.edu.in, likhitha.21bce9931@vitapstudent.ac.in, durga.21bce9938@vitapstudent.ac.in, chaitanya.21bce7607@vitapstudent.ac.in

**ABSTRACT**

As the number and complexity of malware attacks continue to increase, there is an urgent need for effective malware detection systems. While deep learning models are effective at detecting malware, they are vulnerable to adversarial attacks. Attacks like this can create malicious files that are resistant to detection, creating a significant cybersecurity risk. Recent research has seen the development of several adversarial attack and response approaches aiming at strengthening deep learning models' resilience to such attacks. This survey study offers an in-depth look at current research in adversarial attack and defensive strategies for malware classification in cybersecurity. The methods are classified into four categories: generative models, feature-based approaches, ensemble methods, and hybrid tactics. The article outlines cutting-edge procedures within each area, assessing their benefits and drawbacks. Each topic presents cutting-edge approaches and explores their advantages and disadvantages. In addition, the study discusses the datasets and assessment criteria that are often utilized on this subject. Finally, it identifies open research difficulties and suggests future study options. This document is a significant resource for malware categorization and cyber security researchers and practitioners.

## I. INTRODUCTION

### A. HISTORY

Many susceptible devices are now connected to the internet as a result of the development of information and communication technologies like cloud computing and the Internet of Things, which makes them simple targets for attackers. This intentional exploitation has resulted in significant security challenges. The globe has been significantly impacted by the COVID-19 pandemic [13], which has led to a rise in cybercrime and cyberattacks. In recent years, the Internet of Things (IoT) has grown at a remarkable rate [14]. However, due to its diverse application areas and lack of security protocols, IoT has become a prime target for attackers. Various types of cyber-attacks exist, including but not limited to malware attacks, Denial of Service, phishing, and social engineering. According to IBM's [15] cyber threat report, malware has become a significant threat to cyber security and is commonly used by attackers to disrupt the normal operation of systems such as personal computers, servers, mobile and IoT devices [16][17]. Malware is developed intentionally by attackers to achieve their malicious purposes, which includes unauthorized access, leakage of confidential information, system damage, blackmailing, etc. [18]. Based on its harmful properties, malware may be categorized into many types such as viruses, worms, trojans, ransomware, downloaders, and rogue software. Cybersecurity is seriously threatened by the ever-rising quantity of malware instances [19]. Thus, in order to stop malware from inflicting significant harm, it is imperative to find and examine malware instances.

### B. THE DEFENSE-ATTACK-ENHANCED-DEFENSE PROCESS

The Defense Attack Enhanced Defense Process is a cycle of iterative steps taken to improve the security of a system against cyber threats. The process begins with a defensive strategy aimed at preventing attacks, followed by an attack phase that simulates real-world scenarios to identify vulnerabilities. In the enhanced defense phase, the weaknesses discovered in the previous stage are addressed, and the security measures are strengthened. The process then starts again, with each iteration aiming to further enhance the security of the system. The cycle is repeated continuously to keep up with the evolving threat landscape and to ensure that the system remains resilient against new and emerging cyber threats. The Defense Attack Enhanced Defense Process is a valuable approach to ensuring the security of complex systems and is widely adopted in the field of cyber security.

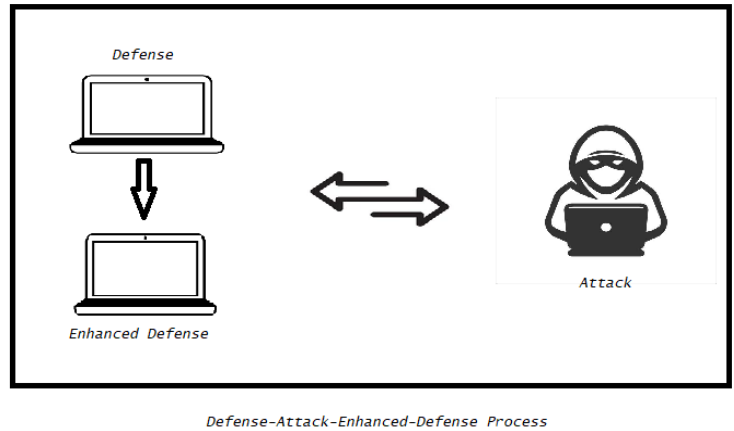

Defense-Attack-Enhanced-Defense Process

**C. MAIN CONTRIBUTION**

The main contributions of our paper are as follows:

- We have methodically examined the Defense Attack Enhanced Defense framework for classifying malware. This comprehensive survey covers machine learning-based approaches for malware classification, methods for adversarial attacks, and defense strategies aimed at improving resilience against adversarial challenges.
- We recognized the major challenges in the Defense Attack Enhanced Defense process and recommended possible future research directions.

**D. RELATED WORK**

In the study by Suciu et al. (2019) [1], a dataset comprising 7,200 benign and malicious PE files was utilized to train and test four distinct malware classifiers: a decision tree, a random forest, a support vector machine, and a convolutional neural network. Adversarial examples were generated using the Fast Gradient Sign Method, and their efficacy in bypassing the detection of the four classifiers was assessed. Demetrio et al. (2021) [2] introduced a novel approach for crafting adversarial malware samples that maintain the original functionality of the malware. Leveraging black-box optimization, their method combines a genetic algorithm and a local search algorithm to identify optimal perturbations to the malware sample. Evaluation on a Windows malware dataset demonstrated the method's effectiveness in generating adversarial samples that elude detection by state-of-the-art anti-malware systems while preserving the malware's original functionality. This approach holds significant promise for advancing the development of more robust anti-malware systems.

The authors [3] applied their framework to the Malimg and NSL-KDD datasets, commonly used in malware detection research. The Malimg dataset comprised 9,427 malware images and 2,784 benign images, while the NSL-KDD dataset included 125,973 instances of network traffic with 22 types of attacks. Through extensive experiments, the authors demonstrated that their framework outperformed several state-of-the-art approaches in terms of accuracy and robustness against adversarial attacks.

Yuan et al.'s paper [4] proposes a black-box adversarial attack method designed to evade the detection of deep learning-based malware classifiers. Using a generative adversarial network (GAN), the method generates perturbations on input malware binaries, creating adversarial examples for the target classifier. Trained with a dataset of clean malware binaries and corresponding labels, the GAN-based approach was evaluated on publicly available datasets (Malimg, Malware, and Drebin). Comparative analysis with other state-of-the-art adversarial attack methods, including FGSM and Deep Fool, revealed the method's efficacy in effectively evading classifier detection while maintaining a low detection rate by the classifiers. The methodology of Kucuk and Yan's [5] paper involves generating adversarial examples for malware classification systems, with the goal of inducing misclassification of specific samples. The authors use a targeted approach, where the adversarial example is designed to be misclassified as a specific class, rather than any incorrect class. They apply this approach to a range of machine learning-based classifiers for Portable Executable (PE) malware, including decision trees, random forests, and support vector machines (SVMs). The authors use two datasets, one consisting of over 2,000 benign and malicious PE files and another consisting of 10,000 benign and malicious PE files. The effectiveness of the adversarial examples is evaluated using metrics such as accuracy, false positive rate, and false negative rate. The methodology presented in the paper [6] is to use an adversarial deep ensemble (ADE) to generate evasion attacks and defenses for malware detection. The ADE is composed of multiple deep learning models, each trained on a different subset of the dataset. The ADE is then used to generate adversarial examples that can evade detection by other deep learning models. The paper introduces two types of attacks: targeted and non-targeted. In targeted attacks, the attacker aims to change the classification of a specific input, while in non-targeted attacks, the attacker aims to misclassify the input as any other class. The authors also propose a defense mechanism that utilizes the ensemble to detect and remove adversarial examples. The datasets used in the experiments include the Malimg dataset, the Microsoft malware classification challenge dataset, and the EMBER dataset.

The researchers in [7] initiated their study with a dataset featuring 9,471 benign and 9,753 malicious executable files. These files underwent preprocessing to extract their opcode sequences using the "IDA Pro" tool, representing low-level instructions executed by a computer processor. Two distinct methods were introduced to create adversarial examples, modifying malicious files to elude detection by deep learning models. The first method employed gradient descent optimization, while the second utilized a genetic algorithm. Adversarial examples were crafted by altering or adding opcodes to the original files while preserving their functionality. For evaluation, a Convolutional Neural Network (CNN) was trained on the original dataset, and its performance was assessed on both the original and adversarial datasets. In [8], the objective was to enhance the interpretability of malware classifiers by generating adversarial examples that shed light on the decision-making processes of these classifiers. Two datasets were employed, one containing 4,566 benign and 4,566 malicious Android APKs, and the other featuring 3,333 benign and 3,333 malicious Windows Portable Executable (PE) files. The authors proposed a black-box attack approach, utilizing a pre-trained malware classifier as a black-box model to generate adversarial examples without knowledge of the model's internal structure. The Fast Gradient Sign Method (FGSM) was employed to generate adversarial examples by perturbing input files based on the gradients of the loss function concerning the input features. Additionally, they introduced a novel hybrid attack method that combined FGSM with a genetic algorithm to enhance the quality of adversarial examples.

In their study [9], the authors employed a Deep Convolutional Neural Network (DCNN) as their malware detection model, training it with a mix of benign and malicious samples. To enhance the model's resistance to adversarial attacks, they incorporated training with adversarial examples generated through the FGSM and BIM attack methods. The training process focused on minimizing the cross-entropy loss function using the Adam optimizer, implementing early stopping based on a validation set. Evaluation of the model's performance included a separate test set of benign and malicious samples, along with a set of adversarial examples generated using the FGSM and BIM attack methods.

Addressing intrusion detection systems (IDS), the authors in [10] introduced a fresh perspective on generating adversarial attacks using Generative Adversarial Networks (GANs). Comprising a generator and a discriminator, GANs were trained to produce attack samples that could evade IDS detection, while the discriminator learned to differentiate between attack and benign samples. Their approach, evaluated on the NSL-KDD dataset, demonstrated higher success rates in generating adversarial attacks capable of eluding IDS detection, all while maintaining a low false positive rate. The authors in [11] explored dynamic analysis-based malware detection systems, which observe program behavior during execution to identify malicious activity. Leveraging the DREBIN dataset for training, a neural network model was developed for malware detection. Adversarial examples were generated using a gradient-based optimization method that maximizes the loss function of the detection model while minimizing alterations to the original input. Additionally, they proposed a novel defense mechanism utilizing a dynamic analysis-based sandbox to assess program behavior and detect any malicious activity overlooked by the neural network model.

The ADE [12] comprises multiple deep learning models, each trained on distinct subsets of the dataset. This ensemble is then employed to generate adversarial examples designed to elude detection by other deep learning models. The paper introduces two categories of attacks: targeted, where the goal is to alter the classification of a specific input, and non-targeted, where the objective is to misclassify the input as any other class. The authors also put forth a defense mechanism utilizing the ensemble to identify and eliminate adversarial examples. Experimental evaluations utilized datasets such as the Malimg dataset, the Microsoft malware classification challenge dataset, and the EMBER dataset.

**E. THE PAPER'S STRUCTURE**

The format of this paper, the unified approach for malware classification is presented in Section II. Section III delves into several malware categorization algorithms that rely on static or dynamic properties. Sections IV and V explain numerous adversarial attack and defense strategies used in the classification of hostile malware. Section VI discusses the difficulties encountered during the defense-attack-enhanced-defense sequence. Section VII investigates potential future study directions and concludes with remarks.

## II. UNIFIED MALWARE CLASSIFICATION FRAMEWORK

The Unified Malware Classification Framework encapsulates the fundamental stages of machine learning-based malware classification methods. It encompasses five primary phases and is applicable for detailing both adversarial attack and fortified defence methods. Adversarial attackers leverage insights garnered from various phases to create adversarial examples, while defenders strive to bolster the adversarial resilience of machine learning-based malware classifiers across the distinct phases of the framework. The five principal phases within the framework are:

- The first phase of the framework is data collection and preprocessing, which involves the gathering of relevant data and the preparation of the data for analysis.
- The second phase is feature extraction, where the relevant features of the collected data are identified and extracted.
- The third phase is feature selection and dimensionality reduction, which aims to reduce the complexity of the feature space and select the most relevant features for the classification task.
- The fourth phase is model training and validation, where the selected features are used to train a machine learning model, and the model is validated on a separate set of data to evaluate its performance.
- Finally, the fifth phase is model deployment and evaluation, where the trained model is used to classify new samples and evaluated based on its performance.

## III. MALWARE CLASSIFICATION ON THE BASIS OF ML

ML-based malware classification refers to the use of machine learning (ML) algorithms to classify malware samples into different categories. In this approach, a dataset of malware samples is first collected and pre-processed to extract relevant information. This information is then used to generate features that can be used to train an ML model. The ML model learns to identify patterns in the features that correspond to different categories of malware. Once trained, the model can be used to classify new malware samples.

- **Binary classification:** In the realm of malware, binary classification involves categorizing malware into two distinct groups - either malicious or benign. Malware, characterized by software intentionally designed to harm or exploit computer systems or networks, undergoes scrutiny to determine its nature. Machine learning algorithms, trained on extensive datasets containing known instances of malware and benign software, are employed to create models capable of accurately classifying files or code as malicious or benign. This facilitates the rapid identification and response to potential threats, forming a crucial element in various cybersecurity systems.
- **Multi-classification:** In the context of malware, multi-classification extends beyond the binary distinction by categorizing malware into multiple classes or families, considering diverse characteristics beyond mere malicious or non-malicious labels. Given that different malware families exhibit distinct attack patterns and behaviors, accurate identification becomes imperative for effective defense. Multi-classification approaches in malware leverage machine learning algorithms like decision trees, support vector machines, and neural networks to classify malware into various families. The feature extraction and selection process in multi-classification is typically more intricate than in binary classification, demanding the identification of features that can differentiate between multiple malware families.

Static and dynamic malware classification are two approaches for detecting and categorizing malicious software.

Static malware analysis refers to the examination of malware without executing it. The static analysis technique involves inspecting the binary or source code of the malware and extracting features such as strings, function calls, and API usage. These extracted features are then used to identify the malware and categorize it into different classes.

Dynamic malware analysis, on the other hand, involves running the malware in a controlled environment to observe its behaviour. Dynamic analysis techniques include monitoring system calls, file system activity, network traffic, and memory usage during the execution of malware. These dynamic features are then used to classify the malware.

Both static and dynamic analysis have their advantages and limitations. Static analysis is faster and less resource-intensive but may miss malware that uses advanced evasion techniques. Dynamic analysis is more comprehensive but requires more resources and may not detect malware that only activates in certain conditions. Therefore, combining both techniques can provide a more robust and effective malware detection system.

TABLE III
COMPARISON AMONG ML-BASED MALWARE CLASSIFICATION METHODS

| TYPES | METHOD | KEY THEORIES | APPLICATION SCENARIOS |
|---|---|---|---|
| Static Methods | Image-Based | Convert binary code features into visual features | Static methods do not need to execute the input examples but are susceptible to obfuscation techniques |
| | Raw Byte-Based | Directly use binary codes as input features to pre serve semantic information. | |
| | Opcode-Based | Opcodes are machine language instructions that contain features of s/w operations. | |
| | Function Call-Based | Functions calls represent the behavioral features of software examples. | |
| | Hybrid Static | Use a mix of features for higher performance | |
| Hybrid Methods | Hybrid Feature-Based | Use both static and dynamic features | Take advantage of both ones |
| Dynamic Method | Hybrid Dynamic | Use a mix of features for higher performance. | Dynamic methods represent the behavioral features but lead to higher overhead. |
| | Hardware-Based | Monitor programs with additional hardware and col lect behavioral features. | |
| | Behavioral Graph-Based | Transform audit data into graph-structured data to extract behavioral features. | |
| | API-Based | Monitor API calls to capture behavioral features | |
| | Network-Based | Monitor network traffic for malicious behaviors such as downloading files and leaking | |

### A. MALWARE CATEGORIZATION BASED ON STATIC CHARACTERISTICS

Malware classification based on static features is a common approach used in the field of cybersecurity to identify and categorize malicious software. Static analysis involves examining the code and other properties of a file without actually running it, which allows for the detection of malicious behaviour before it can cause harm.

Some of the static features that can be used for malware classification include:

1. **File size:** The size of the file can sometimes be an indicator of whether it is malicious or not. Malware authors often compress their code to make it smaller and more difficult to detect.
2. **File type:** The type of file, such as executable, script, or document, can also be an indicator of whether it is malicious or not. For example, executable files (.exe) are often used to spread malware.
3. **File header:** The header of a file contains information about the file format and can be used to identify the file type and other properties.
4. **String analysis:** Static analysis can also involve searching for specific strings of code or text within a file that are associated with known malware families.
5. **Code analysis:** Code analysis involves examining the actual code within a file to identify specific functions or behavior that are associated with malware.
6. **API calls**: Malware often uses specific API calls to perform certain actions, such as accessing the registry or creating a network connection. Analyzing the API calls used by a file can be an effective way to identify malware.

*By analysing these static features, machine learning algorithms can be trained to classify files as either benign or malicious. This approach is particularly useful for identifying new and unknown threats that have not yet been added to antivirus signature databases.*

## B. MALWARE CATEGORIZATION BASED ON DYNAMIC CHARACTERISTICS

Malware classification based on dynamic features is another approach used in the field of cybersecurity to identify and categorize malicious software. Unlike static analysis, dynamic analysis involves executing the malware in a controlled environment and observing its behaviour in real-time. This allows for the detection of malicious behaviour that may not be evident through static analysis alone.

Some of the dynamic features that can be used for malware classification include:

1. **System calls:** When malware executes, it makes various system calls to perform different actions. Analyzing the system calls made by the malware can provide insight into its behavior and help identify malicious activity.
2. **Network activity**: Malware often communicates with remote servers to download additional components or exfiltrate data. Analyzing the network activity of the malware can help identify suspicious communication patterns.
3. **File system activity:** Malware may create or modify files on the infected system. Analyzing the file system activity of the malware can help identify files that are created or modified as part of malicious behavior.
4. **Registry activity:** Malware may modify the registry to achieve persistence or hide its presence on the infected system. Analyzing the registry activity of the malware can help identify suspicious modifications.
5. **Memory analysis:** Malware may use various techniques to evade detection, such as encrypting its code or using anti-debugging techniques. Analyzing the memory of the infected system can help identify such evasion techniques.

*By analyzing these dynamic features, machine learning algorithms can be trained to classify files as either benign or malicious. This approach is particularly useful for identifying new and unknown threats that have not yet been added to antivirus signature databases. However, dynamic analysis can be more resource-intensive than static analysis and may not be suitable for large-scale analysis of large malware datasets.*

**TABLE I. Widely Employed Data Sources for Malware Categorization**

| DATA SOURCE | FILE TYPE |
| --- | --- |
| Virus Share | PE (Portable Executable) |
| Malware Bazaar | PE, Office files, PDF files, scripts |
| VX Vault | PE, Scripts |
| Hybrid Analysis | PE, Office files, PDF files, scripts |
| Virus Bay | PE, scripts |
| Mal Share | PE, Office files, PDF files, scripts |
| The Zoo | PE, scripts |
| Any Run | PE, Office files, PDF files, scripts |
| Open Malware | PE, scripts |
| MISP | PE, Office files, PDF files, scripts |

## IV. ADVERSARIAL ATTACK METHODS

Adversarial attacks involve strategies aimed at manipulating or tricking machine learning models by introducing meticulously crafted inputs, leading the model to generate incorrect or unintended outputs. The prevalence of these attacks is on the rise, particularly as machine learning models find broader applications in fields such as cybersecurity, natural language processing, and computer vision.

### A. ADVERSARIAL ATTACKS ON ML MODELS

1. **"Exploring the Vulnerability of Deep Learning-based Object Detection to Adversarial Attacks" by Xie et al. (2017):** This paper [20] demonstrates how white-box attacks can be used to create adversarial examples that can fool object detection models. The authors propose a new attack algorithm that is specifically designed for object detection models and demonstrate its effectiveness on several state-of-the-art object detection models.
2. **"Adversarial Attacks on Neural Networks for Graph Data" by Zügner et al. (2018):** This paper [21] shows how white-box attacks can be applied to graph neural networks (GNNs) and demonstrates the effectiveness of several attack algorithms on GNNs trained on different graph datasets.
3. **"Adversarial Examples for Semantic Segmentation and Object Detection" by Zhang et al. (2019):** This paper [22] demonstrates how white-box attacks can be used to generate adversarial examples that can fool semantic segmentation and object detection models. The authors propose a new attack algorithm that takes into account the structure of the models and the characteristics of the datasets to generate more effective adversarial examples.
4. **"Transferable Adversarial Attacks for Image and Speech Recognition Systems" by Narodytska et al. (2017):** This paper [23] demonstrates how white-box attacks can be used to create transferable adversarial examples that can fool multiple machine learning models. The authors propose a new attack algorithm that can generate adversarial examples that can transfer between different image and speech recognition models.
5. **"Generating Adversarial Examples with Adversarial Networks" by Goodfellow et al. (2014**): This paper [24] proposes a new approach to generating adversarial examples using adversarial networks. The authors demonstrate the effectiveness of their approach on several machine learning models and show that their method can generate adversarial examples that are more difficult to detect than other white-box attack methods.

*Overall, these papers demonstrate the effectiveness of white-box attacks on a wide range of machine learning models and datasets. They also highlight the need for more robust and effective defenses against adversarial attacks, especially for models that are vulnerable to white-box attacks.*

### B. BLACK-BOX ADVERSARIAL ATTACKS

1. **"Practical Black-Box Attacks against Machine Learning" by Papernot et al. (2017):** This paper [25] proposes a new approach to black-box attacks that is based on the use of substitute models. The authors demonstrate the effectiveness of their approach on several machine learning models and show that their method can generate adversarial examples that are transferable across different models.
2. **"Black-box Adversarial Attacks with Limited Queries and Information" by Ilyas et al. (2018):** This paper [26] proposes a new approach to black-box attacks that is based on the use of a limited number of queries and partial knowledge about the target model. The authors demonstrate the effectiveness of their approach on several machine learning models and show that their method can generate high-quality adversarial examples with limited resources.
3. **"Universal Adversarial Perturbations" by Moosavi-Dezfooli et al. (2017):** In this study [27], a novel strategy for black-box attacks is introduced, leveraging universal adversarial perturbations. The authors illustrate the efficacy of their approach across various image classification models, highlighting its capability to generate perturbations that seamlessly transfer between different models.
4. **"Query-Efficient Black-Box Attack: An Optimization-based Approach" by Chen et al. (2019):** This paper [28] proposes a new optimization-based approach to black-box attacks that is based on the use of a limited number of queries. The authors demonstrate the effectiveness of their approach on several machine learning models and show that their method can generate high-quality adversarial examples with a small number of queries.
5. **"Zeroth-Order Black-Box Adversarial Perturbations for Deep Neural Networks" by Cheng et al. (2019):** This paper [29] proposes a new approach to black-box attacks that is based on the use of zeroth-order optimization methods. The authors demonstrate the effectiveness of their approach on several machine learning models and show that their method can generate high-quality adversarial examples with limited information about the target model.

*Overall, these papers demonstrate the effectiveness of black-box attacks on a wide range of machine learning models and highlight the need for more robust and effective defences against adversarial attacks, especially for models that are vulnerable to black-box attacks.*

## V. STRATEGIES TO STRENGTHEN ROBUSTNESS AGAINST ADVERSARIAL ATTACKS

Various defence mechanisms can be implemented to fortify the adversarial robustness of machine learning-based malware classifiers. Here is a comparison of some of the commonly used defence methods:

- **Adversarial Training:** This approach entails incorporating adversarial examples into the training process of the machine learning model. While it can notably enhance the model's resistance to adversarial attacks, it may concurrently extend the model's training duration and computational demands.
- **Input Pre-processing:** This method involves modifying the input data before it is fed into the machine learning model. Techniques such as data normalization and input randomization can make it harder for an attacker to generate adversarial examples, but they can also reduce the accuracy of the model on legitimate inputs.
- **Gradient Masking:** This method involves modifying the gradients of the machine learning model to prevent an attacker from calculating the perturbations needed to generate adversarial examples. Gradient masking techniques can be effective at defending against certain types of attacks, but they can also reduce the accuracy of the model on legitimate inputs.
- **Ensemble Methods:** This method involves combining multiple machine learning models to improve the overall robustness of the system. Ensemble methods can be effective at defending against a wide range of attacks, but they can also increase the computational complexity of the system.
- **Feature Selection:** This method involves selecting a subset of features from the input data that are most relevant for malware classification. Feature selection can reduce the number of input dimensions, making it harder for an attacker to generate adversarial examples. However, it can also reduce the accuracy of the model on legitimate inputs if important features are discarded.

Overall, each defence method has its advantages and disadvantages, and the choice of method depends on the specific requirements of the application. For example, adversarial training and ensemble methods are effective at defending against a wide range of attacks, but they can also increase the computational complexity of the system. Input pre-processing and feature selection can be effective at reducing the vulnerability to attacks, but they can also reduce the accuracy of the model on legitimate inputs. Gradient masking can be effective at defending against certain types of attacks, but it can also reduce the model's accuracy on legitimate inputs. Therefore, a combination of these methods may be used to achieve better defence against adversarial attacks while maintaining high accuracy on legitimate inputs.

## VI. HURDLES IN THE DAED PROCESS

The Defence Attack Enhanced Defence (DAED) process involves a continuous cycle of developing defences, discovering new attack methods, and improving the defences to counter the attacks. While this process is important for improving the security of machine learning systems, there are several challenges associated with each stage of the process. Here are some of the main challenges:

- **Defense:** One of the main challenges in developing robust defenses is the lack of understanding of how machine learning models make decisions. This can make it difficult to identify potential vulnerabilities and develop effective defenses. Future research could focus on developing explainable AI techniques to provide more insight into the decision-making process of machine learning models.
- **Attack:** Adversarial attacks are constantly evolving and becoming more sophisticated. Attackers can use a combination of techniques to circumvent existing defenses, making it challenging to keep up with the latest attack methods. As such, future research could focus on developing more comprehensive and adaptive defense mechanisms that can detect and respond to new attack techniques.

- **Enhanced Defense:** One challenge in the enhanced defense stage is that some defense methods can inadvertently create new vulnerabilities or introduce new attack surfaces. For example, adversarial training can lead to overfitting or create new attack surfaces that attackers can exploit. Future research could focus on developing more robust enhanced defense methods that do not introduce new vulnerabilities.
- **Evaluation:** Evaluating the effectiveness of defense methods can be challenging. Adversarial attacks are often specific to the target machine learning model and dataset, making it difficult to generalize the effectiveness of defenses. Future research could focus on developing standardized evaluation techniques and benchmarks to assess the effectiveness of defense methods.
- **Scalability**: Another challenge is the scalability of defense methods. As machine learning models become larger and more complex, it can be challenging to implement defense methods without incurring significant computational costs. Future research could focus on developing more efficient and scalable defense methods that can be applied to large-scale machine learning systems.

In summary, the DAED process is a critical component of improving the security of machine learning systems, but it is not without challenges. Addressing these challenges will require continued research and development of new techniques and methods to enhance the adversarial robustness of machine learning systems.

## VII. FUTURE SCOPE

Several potential avenues for future research exist in the domain of adversarial attacks and malware classification. Here are some possibilities:

- **Exploring Ensemble Methods for Enhanced Robustness:** Investigate the use of ensemble methods, which combine multiple classifiers to enhance overall classification performance. Ensembles may prove effective in detecting and mitigating adversarial attacks on malware classifiers. Future research could delve into various ensemble methods, evaluating their effectiveness in countering adversarial attacks.
- **Examining Diverse Adversarial Attack Types:** While much existing research focuses on perturbation-based attacks in malware classification, other attack types, such as those involving feature removal or modification, remain less explored. Future investigations could assess the impact of these alternative attack types and devise strategies for their mitigation.
- **Developing Robust Defenses Against Adversarial Attacks:** Despite existing defenses against adversarial attacks on malware classifiers, many have limitations and can be circumvented by sophisticated attackers. Future research might concentrate on developing more robust defenses capable of better safeguarding against different attack types.
- **Assessing Generalizability Across Classifiers and Datasets:** Existing research often concentrates on specific classifiers and datasets, leaving questions about the generalizability of findings to other contexts. Future studies could probe into how various types of attacks and defenses generalize across different classifiers and datasets.
- **Evaluating Real-World Impact:** While laboratory experiments provide valuable insights, it is crucial to assess the impact of adversarial attacks and defenses on real-world systems. Future research could investigate the effectiveness of different attacks and defenses in practical, scalable scenarios involving real-world malware classifiers.

## VIII. CONCLUSION

This paper offers an overview of the Defense Attack Enhanced Defense (DAED) process in the realm of ML-based malware classification, shedding light on the security vulnerabilities introduced by machine learning approaches. We present a unified framework for malware classification, reviewing both static and dynamic ML-based methods. Furthermore, we delve into the realms of white-box and black-box adversarial attacks on ML-based malware classifiers, examining diverse defense strategies to fortify these classifiers against such threats. Through our analysis, we pinpoint crucial challenges faced by both attackers and defenders in this field and propose avenues for future research. Our findings underscore the dynamic evolution of strategies employed by attackers and defenders in the ongoing landscape of ML-based malware classification. It is our aspiration that this exploration will inspire further research, fostering the development of more effective and resilient malware classification techniques.